# Oral Billiards

*A Physical Model of Speech*

Elaine Y L Tsiang[†]

Monowave Corporation, Seattle, WA

We propose a physical model of speech to explain its precision and robustness. We begin by reducing the dynamics to the bare minimum of polygonal billiards. The symbolic stability of the billiard trajectories against variations in action and the oral cavity geometry forms the basis for precision and robustness in articulation. This stability survives forcing and dissipation to underpin reliable encoding of the trajectories into acoustic emissions. The kinematics of oral billiards and the cyclical nature of the forcing mechanism engender a grammar of the syllable independent of any language. The symbolic dynamics of oral billiards is rendered nearly maximally observable by their concomitant acoustic emissions. Speech recognition is the set of computations on the sub-maximally informative acoustic observables from which the symbolic dynamics of oral billiards may be inferred.

## I. Introduction

The great paradox of speech is its precision and robustness despite its variability. The solution to this paradox has been elusive enough to elicit assertions that speech has no invariant properties [NonInvariance]. Because of the conflation of speech with language, the precision and robustness have been explicitly or implicitly attributed to linguistic context dependency. There is now experimental evidence for animal recognition of phonetic categories to warrant a strong case for the language neutrality of speech[Mice]. Besides, speech is not unique to language, take for example, hand signing or writing systems. It is only one form of the coding component of language considered as a system of communication. As such, the precision and robustness of speech should have a physical basis.

A musical genre called beatboxing makes this physical basis more apparent. fNMR recordings of beatboxing[Beatboxing] in action vividly demonstrate the pugilistic or *ballistic* character of the dynamics, and the significant constraints imposed by the geometry of the oral tract on the dynamics.

The underlying mechanics are quite simple - the tongue as a projectile traversing the oral cavity, hitting various points of the upper surface and bouncing against the jaw line, like a billiard ball.

Billiards as an abstract model of dynamics is of interest because the model can be used to study various properties of dynamical systems independently of complexities in the actual physical phenomena. We will use 2D polygonal billiards[Billiards], which arguably is the simplest version, non-chaotic, but spotted with singularities, to analyze articulation.

† ter@monowave.com

Section II gives a brief exposition of polygonal billiards and the relevant theorems. We specify the oral polygon in section III as the arena of applying polygonal billiards to the geometry of the oral cavity. Based on this geometry, we define the *polygonal alphabet* in section IV. These "particles" of speech live in a 3D space. A broken symmetry in this space will be used to explain some curious relations between some phonetic categories.

Section V defines the basic unit of oral billiards. It leads the discussion of the subsequent 4 sections to end in a definition of the syllable as emergent from a grammar of oral billiards.

The constraint of orbit stability limits the transitivity of the billiard trajectories. Section VI discusses admissible and not so admissible transitions.

In section VII, we add back the complications of forcing and dissipation with the controlled driving of oral billiards and show that the symbolic invariance survives them to underpin reliable articulation.

To complete speech as a coding system, the billiard trajectories are made observable by driving them in such manners that the concomitant acoustic emissions may be redundantly informative on the kinematics. Section VIII redefines the traditional phonetic dimension of *manner* as dissipative driving.

Section IX expands the polygonal alphabet into the fine-grained categories of the *phthongal alphabet*, which now contains the different vowels. The broken symmetry of the oral polygon leads to a unified 3D schema for the phthongal alphabet. Conjugation of the phthongal alphabet with manner completes the phthongal alphabet into the full phonetic alphabet in a 4D phonetic space. In section X, the kinematics of Oral Billiards, manifest as the full phonetic alphabet, gives rise to *a grammar for the syllable*.

Section XI discusses how the limits of the variations that remain after the phonetic categories give rise to the six *prosodic* dimensions of speech. These may be used to augment the phonetic subspace or serve their intrinsically non-symbolic, musical purpose. This 10D phonetic/prosodic space provides a complete characterization of the observable actions of speech.

Section XII identifies the 4D phonetic subspace of the 10D phonetic/prosodic space as the sub-maximally informative observable targets of speech which are manifest in *categorical perception*[MIO]. From reliable articulation and perception result the precision and robustness of speech.

We conclude with a discussion in section XIII of the implications of this model for some long standing issues in speech.

## II. Polygonal Billiards

Polygonal billiards is about a point particle moving within a given 2D convex polygon in the absence of any force except the boundaries. Hence the particle moves with constant velocity in the interior of the polygon. When the particle collides with a side of the polygon, its direction is specularly reflected, its speed remaining unchanged, thus energy is conserved. The corners of the polygon are considered singularities in this setup, because collision at a corner is undefined. Given a velocity and a starting point for the particle, it would then bounce forever inside the polygon, unless it runs into a corner. As stated, any description of the particle's motion would be independent of the scale of the polygon. Hence also the particle's speed. The space of possible p-sided polygons is then of dimension (2p-4), (p-1) from the inner angles that sum to 2π plus (p-3) from the relative lengths. The orbit flow manifold of the billiard map is the product space of the bounding sides with the inward pointing directions.

If the sides of the polygon are symbolically labeled, then any orbit of the particle can be specified by a symbolic sequence. An orbit in which the particle revisits some point with the same velocity is called a periodic orbit. The symbolic sequence would consist of indefinitely repeating subsequences. Many nice conclusions can be drawn about such orbits in *rational* polygons, whose inner angles are rational multiples of π, but not as much for generic polygons. As our interest in this paper is to apply polygonal billiards to a very messy real-life situation, we cannot restrict ourselves to rational polygons only. Nevertheless, the most important question for our purposes - whether billiard orbits of interest are stable against two kinds of perturbations - can best be explored with periodic orbits, stability being defined as the invariance of the symbolic subsequences. One kind of perturbation is variations in the initial conditions in phase space - the particle's initial location and its direction. The other kind is variations in the geometry in the space of polygons. There is a theorem which says that if a polygon has a periodic orbit, then that orbit is also stable against both kinds of perturbations[1]. So the stability problem becomes a hunt for the existence of periodic orbits. This hunt is more difficult than one would expect, and is not completed for even the simplest of polygons, the triangle.

For our purposes, it suffices to make use of what has been hunted down already for the triangle, even though we will be dealing with an often irrational six-sided and sometimes five-sided polygon. In addition, we can relax our requirement from absolute stability to stability within a reasonable decay time. It has been proven in general that if an orbit is not periodic, it would eventually run arbitrarily close to a corner[Dichotomy]. But if it takes a significant number of collisions before it gets near a corner, we have sufficient stability for our purposes.

In acute triangles, there is a periodic orbit following an inscribed triangle called a Fagnano triangle formed by the intersections of perpendiculars from the vertices to their opposite sides (Fig. 3). Fig. 1 depicts a periodic orbit displaced from the Fagnano orbit, showing that there is a continuum of such orbits. If we label the sides from left clockwise as [b], [c] and [a], then these orbits are represented by the symbolic sequence [bac] and their shifted and reversed versions; illustrating a simple example of symbolic invariance against displacement along the sides.

For right-angled triangles, there also exists a periodic orbit that hits the hypotenuse twice perpendicularly tracing three sides of a rectangle (Fig. 4) repeating the symbolic sequence [babcac].

If one of the angles is very acute, the enclosing two sides are almost parallel, then orbits that hit the sides almost perpendicularly towards the acute angle will run into that corner after many bounces, with a symbolic sequence of [ba] until they hit the corner (Fig. 5).

When we come to add forcing and dissipation to the dynamics, we are ultimately only interested in the stability of the first 2-symbol sequence per forcing. Even before stability, the most important property of polygonal billiards for our purposes is that it is intermittently ballistic. There is necessarily a flight in the interior of the polygon between collisions. The flight ensures the separability of the collisions.

---

1     The construction for the theorem proving stability against geometric variations can also be used to prove stability against kinematic variations, which translates to articulatory variations.

## III. Oral Billiards

First, make the particle in motion a finite-sized billiard ball. This rescales the geometry, but all of polygonal billiards apply[2]. This billiard ball represents an idealized tongue, including the simplification that it is not attached. Fig. 2 shows the geometry of the oral cavity as an open 6-sided polygon. The sides are labeled by their phonetic symbols. Side [θ] is a gross simplification of the complex of the upper lip, teeth, and alveolar ridge in an actual oral cavity. Side [x] is similarly a simplification of the convoluted ending of the soft palate in the velum. Side [ʔ] , which we will call the jaw line, represents the side that the tongue is to be attached to. It opens the polygon into the pharynx, represented by a channel that ends at the glottis, marked by the same symbol. The diamond represents the fixation of the jaw line relative to the lower jaw, represented by a dotted line hinged at a large circle. The up/down mobility of the jaw and its hinging makes it possible to orient the jaw line at a wide range of angles. The shorter dotted line representing the lower teeth does not constitute one more side because the billiard ball, once attached, cannot collide with it.

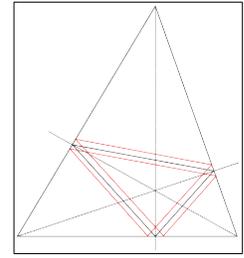

*Fig.1 A periodic orbit*

The scale of this oral polygon is best set by the length of the jaw line. Its length varies on the order of 6 to 10 cm, and remains fixed upon maturity. The scale invariance of polygonal billiards underpins the "speaker independence" of the dynamics of the billiard in the oral polygon.

As wetware, there are no corners in the actual oral cavity. But the roof of this cavity has varying curvatures. The simplification above makes the low-curvature sections into the polygon sides, which then squeezes the high-curvature sections into corners. The resulting sides of the polygon and the corners [ɸ] and [ʔ] happen to coincide with the dominant places of articulation of traditional phonetic categories.

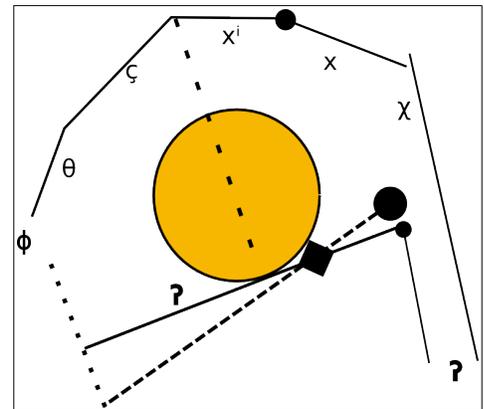

*Fig. 2: The Oral Polygon*

There are 2 movable sides that may change the geometry of the oral polygon. The small circles indicate where they hinge. The jaw line, [ʔ], may be lowered as a whole or rotated down when the jaw hinges down, thereby increasing the inner angle between side [ʔ] and side [χ], The velum is shown slightly open, with side [x] slightly hinged down, thus reducing the angle between it and all other sides except [χ]. When the velum is closed, side [$x^i$] and side [x] become one, reducing the hexagon to a pentagon.

Note, by the same scale invariance, that the size of the billiard ball could be dynamic, in order to approximate the various parts and conformations of the ductile tongue as a ball. When collisions are with the tip of the tongue, the ball would grow, in transit, smaller. When collisions are with the top or towards the root of the tongue, the ball would grow bigger to correspond to the size and curvature of the tongue. This dynamic size of the billiard is equivalent to dynamically changing the scale of the polygon with the billiard in flight. Again, the kinematics are immune to this change of scale.

---

2  For visual clarity, the diagrams of the orbits do not explicitly show the finite size of the billiard ball.

## IV.   The polygonal alphabet

The alphabet of the billiard's collisions in the oral polygon consists of the 6 labeled sides and the corner at the lips. Table 1 lists the corresponding places in the oral cavity. The glottis is not part of the oral polygon, but the action of the billiard hitting the jaw line will be observable only when the glottis or pharynx admits a source of acoustic excitation. Both the glottis and the jaw line are labeled [ʔ] to indicate this relation. The labial place, like the glottal place, is not in itself a site of a billiard collision, except in the limit of the billiard running into (or out of) its corner. More importantly, it stands for the opening of the jaw.

*Table 1. The polygonal alphabet*

| place | symbol |
|---|---|
| glottal | ʔ |
| labial | ɸ |
| dental-alveolar | θ |
| alveolar-palatal | ç |
| Velar-palatal | $x^i$ |
| velar | x |
| uvular-pharyngeal | χ |

The billiard may hit at varying points along a side. All such actions are represented by the same symbol. We will call these 7 symbols the polygonal alphabet.

There are finer phonetic categories than the polygonal alphabet in the International Phonetic Alphabet [IPA] We will eventually construct a full phonetic alphabet[IHA] which is a fine graining of the polygonal alphabet. The members of this full alphabet will more densely annotate the sides of the polygon, as well as the interior proximal to the sides of the polygon, but differing in tongue conformation. This suggests that this alphabet lives in a 2D space. The simplification of the oral polygon, however, evinces a bilateral symmetry that is obscured in an actual oral cavity. The two top right sides of the polygon mirror the two top left sides across he dotted line through the billiard in Fig. 2. This broken symmetry will be central to the schema for organizing the fine-grained alphabet. It is therefore helpful to make this broken symmetry explicit by adding a dimension, albeit one that consists of only three values. We will name one of the values PAL, short for palatal-alveolar-labial, the other VUP, for velar-uvular-pharyngeal. A more intuitive naming would be *front* and *back*. But these two terms have long been used to characterize the so-called vowel space, which is subsumed under the third value of Glottal. To avoid confusion, we adopt the more awkward naming, We will however abuse tradition by naming this dimension Place, when it is more like a coarse graining of its traditional meaning.

## V.   The Syllable

The basic unit of billiard action consists of two boundary collisions (Cs) separated by the ballistic flight (X) between them – denoted CXC. For example, [x]X[ʔ], [ʔ]X[θ] or [θ] X/[x]. Note that Cs occur at points in time, and only Xs have duration. During X, the billiard is in flight in the interior of the oral polygon.

The symbolic description dispenses with the details of X, and is only concerned with the collisions. Because the majority of the stable trajectories, described below, are between the top sides and [ʔ], C[ʔ] and [ʔ]C[^3] actions abound, and in longer orbits, "words" like C[ʔ]C(or [ʔ]C[ʔ]) dominate usage in languages, and are perceived to be the "atoms" of speech, called syllables in traditional phonetics.

---

3   These correspond to semi-syllables in traditional phonetics.

There are a few languages[4], known to have clusters of consonants, that would seem less anomalous if we take CXCs as the atoms of speech.

In the following sections, we will see how the inherent transitivities of oral billiards, dissipative driving of their actions, and the concomitant acoustic emissions that enable their observation, give rise to a grammar for how these "atoms" of speech may be assembled into larger entities. The major non-terminal in this grammar would best be named the syllable.

## VI. Diphthongal Transitivity

Fig. 3 shows a oral billiard symbolic sequence [θʔχ] based on the Fagnano orbit. A similar one, [çʔχ], is obtained with side [ç] that also intersects at an acute angle with [χ].

Fig. 4 Shows an orbit [θʔx] that is based on the side [θ] intersecting at roughly a right angle with side [x]. If the velum hinges down, then [θʔx] will be morphed into a Fagnano orbit.

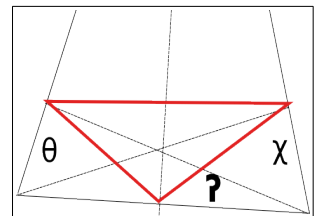

*Fig 3. [θʔχ]*

Fig. 5 shows the almost parallel sides [x$^i$] and [ʔ] bounding the near-perpendicular bounces. Similar bounces can be trapped between sides [ç] and [ʔ] when the jaw line is lowered and hinged down to make these two sides almost parallel. These are the stable orbits that come from acute angles formed by skipping over two or more intervening sides, excepting side [ʔ], of the oral polygon. The corresponding CXCs are the most commonly occurring ones in spoken languages.

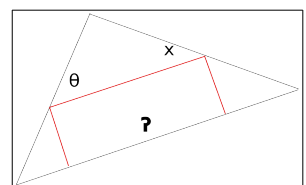

*Fig 4. [θʔx]*

Trajectories that hit sides more adjacent than two tend to be glancing, and unstable with respect to small variations in initial velocity and position. This instability in articulation is acknowledged as "tongue twisters"[5] in nursery rhymes, and may underlie the apparent predominance of one of two adjacent sides in some spoken languages[6]. It is stabilized in practice by injecting a [ʔ] to convert the trajectory into a near-perpendicular bounce.

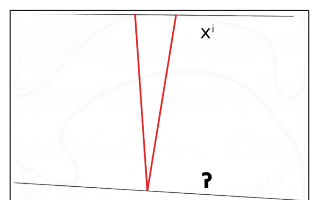

*Fig 5. [x$^i$ʔ]*

Note, to utilize the rectangular orbit in Fig. 4, it would be convenient to slide the billiard along the hypotenuse [ʔ]. From this comes the diphthong, two vowels back to back, of traditional phonetics.

## VII. Dissipation and Forcing

Oral billiards would be a lame game if we cannot drive the symbolic sequences. In addition, speaking obviously detracts from the elastic and specularly reflective motion of polygonal billiards.

---

4   For example, Tashliyht
5   For example, in English, "s", associated with [θ] and "sh", associated with [ç].
6   For example, [χ] predominates over [x] in French. The opposite is true for Germain. [ç] predominates over [x] in Russian.

The obvious way to drive oral billiards is at the collisions by dissipating the entering momentum and forcing the exiting momentum. What remains invariant are the polygon sides of the collisions and the directions of the ballistic flights between them.

In general, adding dissipation to an energy-conserved dynamic system shrinks phase space, but does not alter the topology of the flow manifolds. For oral billiards, dissipation leaves the symbolic orbits intact, but sets an upper limit on the length of the "words".

However, the more efficient the control, that is, the less the dissipation and the more it adheres to the underlying periodic orbits, the more contextually dependent the motion would appear to be. The trajectory is determined for all time from one point in phase space. The more is known about the trajectory, the closer we are to locating any point on the trajectory in phase space. The apparent context dependence is inverted. It is not the dependence of a point in configuration space on a long context, but that the context, that is the complete trajectory, is determined by any one point of the trajectory in phase space.

In fact, the driving impulses are unambiguously separable by the ballistic flight between collisions, and the manners of motor control can make each ballistic flight independent of any context beyond the collisions bookending the flight.

As a spoken language evolves, the people using it become more efficient at it, and tend to omit much of the work of dissipation, and relax into tongue flappings, tappings and bounces.

## VIII. Sound Manners

Oral billiards are made observable by sound, for which the oral cavity is converted into a musical instrument, or, more precisely, a dynamically switched ensemble of musical instruments. Sources of energetic excitation near the open corners or inside the cavity generate the acoustic waves that either radiate directly from open-able corners, such as the mouth or the velum, or probe the cavity as standing waves.

Playing the oral instruments consists of various elaborations of the billiard collisions which emit sounds and dissipate the momentum of the billiard at the same time. Also, changing resonances corresponding to the changing cavity shape caused by the moving billiard ball, leave telling signatures on the emitted sounds. The cavity resonances shape the relative amplitudes of any broad-band acoustic emission to result in peaks and valleys. The peaks are called *formants*, and the valleys are sometimes referred to as *anti-formants*.

The correspondence between the oral cavity shape and the formants has been much studied as the spectral filtering property of the oral cavity as a tract, with occasional extra side tracts from the nasal passage to pockets formed from different conformations of the tongue, approximated to varying degrees by concatenating a few or many tubes. This "tube" model of speech emphasizes the filter effect of the tracts on the excitation sources. For observing oral billiards, we emphasize the *changing* formants, called *formant transitions*, as informative on the ballistic flight between collisions, which is also informative indirectly on those collisions, or symbols, themselves. The variances in the acoustics corresponding to the dynamics are large, the reason for the apparent absence of any invariance. Nevertheless, observables with high variances may still be sub-maximally informative on the targets, namely the polygonal alphabet, as will be discussed in section XII.

In the following, the collisions against the top sides of the oral cavity and the the labial corner will be termed consonants, and the collisions against the jaw line vowels.

For consonants, the collisions themselves are also directly observable by the different ways they are being impulsively driven, formally called *manners i*n phonetics. The *plosive* manner **/P/** stretches the instance of collision into an extended period, during which the back section of the oral cavity is more or less closed off by a seal of the tongue against a side, sometimes accompanied by air rarefaction, and little sound is emitted during this closure. Then the closure is rapidly opened, resulting in a small explosive sound which can be emitted before the section in front of the collision has been fully acoustically coupled to the closed off section. In the bilabial plosive, which is made with the lips, the advanced portion of its little explosive sound is not subject to any cavity filtering.

The *fricative* manner **/F/** also extends a collision by softening the seal, during which air is forced through the constriction to generate noise from turbulence, which, like the little explosions of the plosives, is not all subject to back cavity resonance.

The plosive and fricative can be combined as plosion following immediately by constriction, or constriction followed by closure, referred to as "affricates".

Much the same actions as those for plosives are involved in the *nasal* manner **/M/**, except in this case, the soft palate is hinged down, which changes the oral cavity shape, and opens up the velum to bring in the nasal cavity resonances and corresponding reductions in the oral cavity resonances . The same closure as in the plosive is still the extended collision, but it is accompanied by the nasal resonances and the oral cavity anti-resonances.

The manner that comes closest to an un-dissipated billiard collision is called *approximant /V/* for consonants, and *semivowel* **/H/** *f*or vowels. This manner does not extend the collision, reducing the actions to taps (or flaps for the ductile tongue) and bounces against the jaw line.

The bounce against the jaw line may be extended by holding the billiard against the jaw line during which the oral cavity is *unobstructed*, more or less. This is the manner of *vowel /***A***/***.**

The nasal, approximant/semivowel and vowel manners do not directly emit sounds. They are made observable by a separate energetic excitation, which is often, but not necessarily, provided by vibrating vocal chords. It can, for example, come from a fricative formed simultaneously, or, co-articulated, by constricting the pharynx, or even the glottis[7]. In vowels, the relatively static shape of the oral cavity is made observable solely by the formants. Likewise for the approximant and semivowels. It is the change in the formants corresponding to the flight of the billiard in the interior of the polygon that is informative on this manner.

The distinction between approximants and semivowels is worth elaborating from the perspective of observing the oral billiards. The approximant is that instantaneous configuration of the oral cavity when the billiard ball is proximal to the point of collision with the top sides of the polygon. At the beginning or end of a collision, the front cavity resonances are predominant. At the instant of the approximant, the cavity in front of it and the cavity back of it are acoustically coupled. Hence there may be a *discontinuity* in the acoustic observables of the collision and the approximant at the time resolution of auditory perception, on the order of 5 to 10 ms,. No such discontinuity obtains between the semivowel and its collision, which in this case is a vowel. So the semivowel *approximates* a vowel, but the approximant may be perceptually discontinuous from its collision.

---

7 So-called pharyngealization. An external vibratory source in contact with the throat has the same effect.

In those manners where the billiard collision is extended, the motion may freeze in that interim. This freezing is most obvious in the closure of the plosive manner. Excitation sources can also cease. The motion thereby becomes unobservable. These unobservable moments have been given their own manner of *closure* /‿/, and are *interior* to the collisions. Although equally unobservable, they are logically distinct from the general absence of speech.

Traditional phonetics classifies turbulent excitation at the glottis as a glottal fricative, and considers the glottal stop just another plosive. The glottis is not a site of collision of oral billiards, glottal fricatives are still collisions against the jawline, hence they are relegated to the vowel category. The distinction between vowels and the traditional glottal fricatives becomes prosodic, to be discussed in section XI. Glottal stops mark the fast initiation of vowels just as plosives mark that of fricatives. Glottal stops are therefore a separate manner /ʔ/. As a result, there is not just one glottal stop, but as many as there are vowels. There are, analogously, also as many glottal affricates.

## IX. The Phthongal Alphabet

*Table 2. The phthongal alphabet*

|  | Back' | Back Like' | Central' | Front Like' | Front' | Front | Front Like | Central | Back Like | Back |
|---|---|---|---|---|---|---|---|---|---|---|
|  | PAL | | | | | VUP | | | | |
| close | ɸ | f | θ | ɕ | ç | xⁱ | xᶦ | xᵊ | | xᵘ |
| closeLike |  |  | s | ʃ |  |  |  |  | | xᵒ |
| closeMid |  | ɭ | ʂ | ʎ |  | χᶦ | χᵊ | | | χᵒ |
| mid |  | ɫ | ɹ |  |  |  | | ħᵊ | | ħᵒ |
|  | Glottal | | | | | | | | | |
| close |  |  |  |  | i |  |  |  | | u |
| closeLike |  | ʊ | ï | ʏ | ʏ | ʏ | ï | ʊ |  |
| closeMid |  |  |  |  | e |  | ɘ | | | o |
| mid |  |  |  |  |  |  | ə | | |  |
| openMid |  |  |  |  |  | ɛ | ɜ | | | ɔ |
| openLike |  |  |  |  |  | æ | ɐ | | |  |
| open |  |  |  |  |  |  | a | | | ɑ |

We now come to acknowledge that the tongue is not a detached hard billiard ball, but attached and ductile, and may take on different shapes and present different parts of the tongue to be in contact with the top sides of the oral polygon in any collision. These non-ballistic actions of conformation may modify the acoustic resonances, and are therefore observable. We will call such conformations *phthongs*. In Table 2, we present the conformational elaboration of the polygonal alphabet into the fine-grained *phthongal alphabet*. The symbols are those of fricatives and vowels of traditional phonetics because these two manners have the full complement of phthongs. The tabulation indicates the relative situation of the symbols with respect to the dimensions frontBack and openClose. It is not a direct mapping into the oral polygon, nor does it imply a distance metric.

The fine graining of the glottal Place gives rise to the traditional vowel space of the two dimensions of frontBack and openClose. The symmetry between PAL and VUP should give rise to a duplicate vowel space, the reason for the columns with the primed frontBack labels. The close front vowel [i] is in fact made with the highest point of the tongue closest to the sides /xⁱ/ or /ç/ of the oral polygon. That is, the billiard ball is sitting at the line of broken symmetry between PAL and VUP, but raised up from its relaxed position, which defines the vowel [ə], the shwa. In the position of the shwa, the mouth is slightly open, and the jaw line is positioned for the rectangular orbit of Fig. 4 when the mirror symmetry is least broken, and the cavity between the tongue and the top sides most approximates a uniform tube. Between [i] and [ə], the tongue is still sitting at the line of broken symmetry, but the gap between the tongue and the top sides widens, as the tongue is lowered, tracing out the front/frontLike vowels [ʏ] and [e]. The mirror symmetry forward and backward of this position indicates two ways to articulate the closeLike vowels [ʏ], [ɨ] or [ʊ]. The PAL version of these vowels has a back cavity longer than the front one, and the VUP version is reversed. But the formant structure is degenerate with respect to PAL vs. VUP. As the tongue lowers further without opening the jaw, it is no longer possible to push it forward or backward due to its bulk and its attachment. Past the shwa, the jaw begins hinging down, and there is now room again for the tongue to move forward and backward, realizing the remaining vowel space on the VUP side. Thus the curious "waist" of the vowel space at the shwa.

The symmetry between PAL and VUP is more fully realized for the non-glottal phthongs. The empty cells between PAL and VUP should be read as null, meaning their neighbors are to be read as contiguous. The emphasized phones in deeper-color cells are of the polygonal alphabet, and all can be realized as standalone plosives.

The plosive realization of lighter-color cells on the PAL side are usually affricates. The closeMid and mid consonants [ɭ], [ʂ], [ʎ], [ɬ] and [ɻ] are most familiar as approximants, and are observable due to side-channel resonances and anti-resonanaces from different tongue conformations[8].

The traditional velar has already been split into two sides of the oral polygon, [xⁱ] and [x]. It is now further fine-grained into 5 phthongs. The superscripts on the velar phthongs are not an indication that they are conditioned on an adjacent vowel context, but that the billiard is positioned in frontBack and openClose with acoustic correlates mirroring those of [ç], [θ] and [ɸ]. [xʔ] diphthongs will often transit against the vowel context. For example, a velar to an open backLike vowel is realized as [xᶦ ɑ] without a [iɑ] vowel diphthong, and a velar to a closeLike front vowel as [xᵊʏ] with no [əi] vowel diphthong.

Similarly, the uvular-pharyngeal side has been fine-grained into 5 phthongs, hitting the uvular-phayngeal side at high to low points. The constrictions are made with the back or even the root of the tongue, whereas the corresponding PAL phthongs are usually conformations of the tongue tip or blade. So it is relatively easy to "co-articulate" corresponding PAL and VUP phthongs. The formant structure of these closeMid and mid phthongs, like that of the vowels, is degenerate with respect to PAL and VUP. A non-native speaker may mistake a PAL phthong for its native VUP counterpart, e.g., [ɭ] for [χᵒ].

---

8   Lateral articulation and retroflexion.

Note that the phthongs derive their symbolic stability only from the underlying polygonal alphabet. The vowels, for example, being all fine grainings of the glottal Place, have no inherent symbolic stability among themselves. Only the most extremal vowels, [i], [u] and [a]/[ɑ], are easily distinguishable one from another.

The phonetic alphabet is obtained when the phthongal alphabet is conjugated with the different manners: Glottal with /A/,/ɑ/ and /H/ and PAL and VUP with /F/, /P/ and /V/, and all phthongs with /␣/. They live in the 4D space of $FrontBack \times OpenClose \times Place \times Manner$. A full recitation is to be found in [IHA].

## X. A Grammar for the Syllable

Some organizing scheme is needed to regulate the muscular actions that generate sequences of phones, that is, articulation. The main constraint is the synchronization of the muscular actions with the mechanism for marshaling the energy needed to drive the billiard, that is, breathing. This calls for a grammar whose major non-terminal is cyclical, and has a duration on the order of a breath cycle. The decoding of such phone sequences would also rely on this grammar. The basis for this grammar is the kinematics of oral billiards.

Fig. 6 presents a grammar with the syllable as the major non-terminal. The relations of the syllable to the onset and rhyme, and the constituents of rhyme as being of nucleus and coda are from traditional phonological characterization of the syllable. The rest of our formulation departs from but does not contradict this traditional characterization.

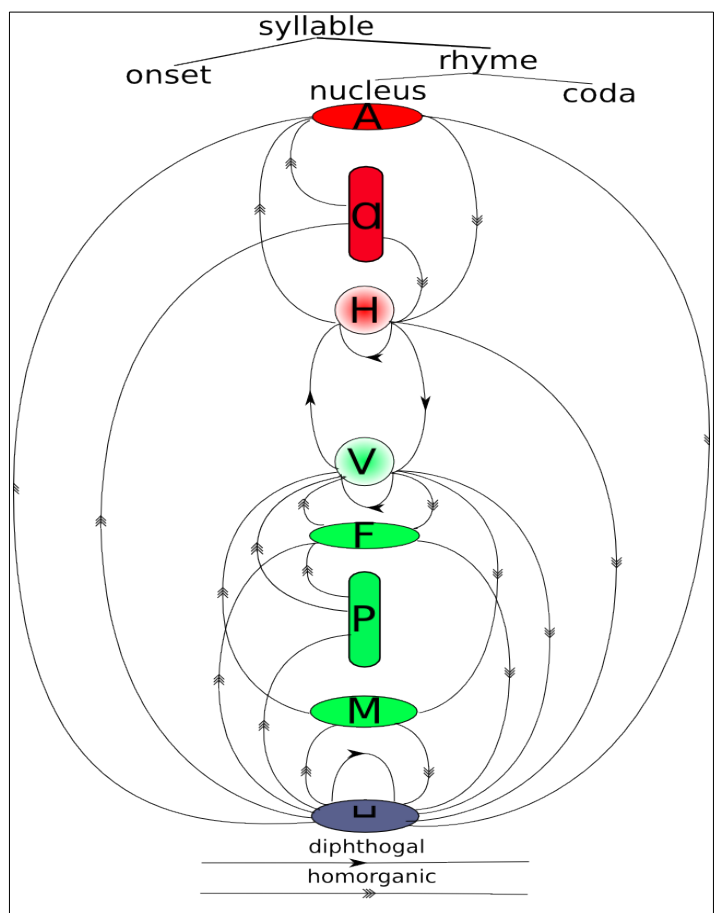

*Fig. 6  A grammar for the syllable*

The nucleus admits any of the manners with nodes represented by various circles, ovals and bars, except for closures. The onset admits any upwardly directed arc on the left side, and the coda admits any downwardly directed arc on the right side. A syllable boundary results at a node when reached with a down arc, and departed from with an up arc. The vowels may not be a syllable boundary because only up arcs enter and down arcs exit from this node. /ɑ/ or /P/ also may not be a syllable boundary because they have no entering down arcs. We can also define a syllable center, or a *syllabic*, at any of the nodes when reached with an up arc, and departed from with a down arc. Only /P/ and /␣/ may not be syllabics.

The ovals represent manners that admit self transitions of the same phthong, that is, they are *continuants,* including the closures. The bars and circles, that is, /Cl/, /P/, /H/ and /V/ do not admit such self transitions.

Both the up arcs and the down arcs that have double arrows are labeled as homorganic, meaning they are of the same phthong. Those arcs exiting or entering node /⌣/ represent a phase change inside a collision. The arcs departing a /V/ or /H/ node represent entries into a collision, whereas those entering a /V/ or /H/ node represent exits from a collision.

The arcs labeled diphthongal, two between /H/ and /V/ and three self-loops of /H/, /V/ and /⌣/, represent the only admissible transitions between different phthongs, that is, flights inside the oral polygon or slides along the jaw line.

## XI. Prosody

The remaining dimensions of speech as captured in the acoustic emission are prosodic. Some delimit the actually realized invariance. The scale invariance of the oral billiards in actually realized oral tracts is limited to roughly 6 cm to 10 cm in the size of the lower jaw. Oral billiards in a tiny oral cavity or a gigantic one would exhibit the same kinds of properties, but would not be perceived as normal, although within narrow deviations, perceivable as speech. In addition, due to the specially evolved descended larynx, which humans can lower and raise, protrusion of lips and tightening of facial muscles, the effective tract length[9] for acoustic emissions can be varied while speaking.

We hear speech over a range of frequency spanning roughly 60 semitones (5 octaves) out of a total of about 8 octaves of the human auditory range. The relevant range for any given speaker is less than 5 octaves, on the order of 4 octaves around a frequency proportional to the inverse of the scale of the oral cavity, which we will call the pin. The pin varies across speakers on the order of two thirds of an octave (corresponding to the range of jaw sizes), and varies dynamically over about 1/3 octave in individual speakers from varying the length of the vocal tract as described above. In traditional phonetics, this modest dynamic variance is termed rounding when the oral tract is extended, and fronting when it is contracted from its relaxed length. Note fronting is really more tongue centering near the line of broken symmetry. Fronting may be accomplished by either facial tightening to shorten the the PAL side, or by raising the larynx to shorten the VUP side; and rounding by lip protrusion or lowering the larynx. We will use the term rounding to refer to all these tract length variations. Rounding is exploited for lexical contrast in some languages.

We hear sounds, therefore speech, over a range of overall loudness spanning roughly 96dbs(16octaves) above a threshold. Like frequency, the necessary dynamic range of speech is less than the full range, on the order of 30dbs (5 octaves) in easy circumstances, but can be less than 0 db in noise, indicating that human listeners hunt down temporary patches of frequency with higher local SNR[10]. These may still be as low as 1-3 db. The variation in this local SNR over the duration of a few syllables is involved in the perception of emphasis or stress.

---

9   From the glottis to the lips.
10  Called *glimpses* in perceptual studies of speech.

We hear speech over a range of speaking rates spanning 4 octaves, from around 2 cps to around 32 cps. The cycle being that of a billiard action of CXC. There are stringent time limits to most CXC actions, except those involving continuants. These may be almost indefinitely extended, or shortened to flaps, taps or bounces. A significant portion of the 4 octaves are taken by the variation in continuant durations, which is the other important factor for the perception of stress.

In addition, simultaneously with hearing speech, we are aware of whether the vocal cords are in vibration, called voicing in traditional phonetics, and if vibrating, its frequency, or pitch. The most extreme realization of voiced speech is singing. The perception of voiced speech is in fact independent of vocal chord vibration. It can be from any source of acoustic excitation along the pharynx down to the glottis, which is the case, for example, in whispered speech. The important observable is the time of onset of the acoustic excitation from the nucleus of the syllable.

Because the velum can be raised or lowered if not precluded by other actions, nasality is not solely phonetic.

Each of these 6 non-symbolic dimensions, rounding, loudness, speaking rate (or duration), voicing, pitch and nasality, can be used to augment the symbolic space. For example, the voicing dimension is most reliable as a binary variable, and is almost universally utilized as such in spoken languages. In some languages[11], voicing has multiple levels for lexical contrast. Binary nasalization is also sufficiently reliable to create a duplicate subspace of nasalized vowels, albeit conditioned by nearby nasals. Similarly for rounding, as mentioned above. Pitch, called tone in traditional phonetics, can be utilized up to 7 levels and over two changes of phase, for lexical contrast.

In general, the prosodic dimensions are utilized for non-symbolic or musical effects, for which they are named. They are more elementary than the traditional phonological constructs, such as stress, rhythm, tone levels or contours. They could be used to redefine the traditional terms, or to define new finer-grained phonological variables.

## XII. Sub-maximally Informative Observables

Considering speech recognition as an inference problem, the symbols residing in the 4D phonetic subspace of oral billiards constitute the target observables to be inferred via acoustic emissions. We have shown that the target observables are symbolically invariant against changes in the geometry of the oral cavity and imprecision in the articulatory maneuvers. The inference must then match up the acoustic observables to the target observables so as to give an accurate estimate of the targets. We have seen that the sequencing of the target observables is governed by a grammar. This grammar would endow the acoustic observables with the same regularity to facilitate their discovery in the general open-channel acoustic environment. The experimental phenomenon of categorical perception indicates that the targets are nearly maximally observable, or sub-maximally observable [MIO] from the acoustic emissions. The problem of speech perception is then to discern what these observables may be, and the set of computations on them from which the symbolic dynamics of oral billiards may be inferred.

---

11  For example, Hindi.

## XIII. Conclusions

**Speech is Special for Language**

Oral billiards imply that speaking, like hand signing, or playing a musical instrument, or a sport, is a controlled sequence of gestures. From the perspective of generating the gestural sequences, they would require similar neuro-motor actions and co-ordination. In this sense, the evolution of speaking is on a continuum with the evolution of other human motor capabilities, such as bipedal locomotion or projectile throwing.

Oral billiards, however, unlike the other motor capabilities, including hand signing, are inherently symbolic, as described in this paper. We conjecture that the evolution of speech involved the geometry of the whole oral apparatus, not just the descent of the larynx. Further, the skeletal, muscular and neural formations around the oral tract that drive the oral billiards have exapted to render the kinematics nearly maximally observable by their acoustic emissions. On the perceiving end, the auditory components of the human brain have co-evolved to reliably extract the symbolic sequences from the acoustic environment. Hence speech is uniquely suited to be the phylogenetic and ontogenetic enabler of language.

**Long Live the Segment**

Speech is indeed like beads on strings, and not scrambled eggs. Successful decoding means the beads, called segments, have been extracted, and the strings, the billiard trajectories, which constitute the bulk of the lower-level computations, are then garbage-collected, leaving the higher-level mind with the illusion that the beads, seemingly abutting and covering, are all there ever were.

**The music of different languages**

The grammar of the syllable derived from oral billiards is language-neutral. The first linguistic level up from the phonetic, the phonological, pertains to the restriction of the language-neutral syllable to a particular language. This restriction is accompanied by a generous mix of voicing, nasality, rounding, pitch, loudness and duration. The first perceptible distinction between spoken languages is, like that between dialects within a language, prosodic, that is, musical.

**A mechanical vocoder based on oral billiards**

Current material and fabrication technologies should be sufficient to realize a mechanical model of oral billiards. Some of the non-trivial engineering problems glossed over in the above theoretical discussion, such as how best to attach the billiard, have to be solved. Viewing the design of such a physical model as that of a musical instrument could be helpful. Such a model is capable of falsifying, or limiting, the claims of the thesis that the invariance of speech is derived from the properties of oral billiards.

Such an experimental model would also be a possibly cheaper alternative way to directly study the acoustic correlates of speech articulation because one can exactly control the mechanical action without using a human intermediary.

**Machine Speech Recognition based on oral Billiards**

We have developed a computational speech recognition system based on Oral Billiards[12]. The system outputs sequences of the full IHA phonetic alphabet admissible under the non-linguistic oral billiards grammar of the syllable as marks in time diacriticized with the 6 prosodic dimensions, thus providing a universal high-resolution phonetic-prosodic decoder for all spoken languages.

---

12  To be published.